\documentclass[iop,apj]{emulateapj}

\pdfoutput=1 %for arXiv submission
\usepackage{amsmath,amstext}
\usepackage[T1]{fontenc}
\usepackage{apjfonts} 
\usepackage{threeparttable}
\usepackage{hhline}
\usepackage{color}

\usepackage{hyperref}

\shorttitle{Large SFE variation in NGC\,2276}
\shortauthors{Tomi\v{c}i\'c N. et al.}

\begin{document}

\title{Two Orders of Magnitude Variation in the Star Formation Efficiency Across the Pre-Merger Galaxy NGC 2276}

\author{Neven Tomi\v{c}i\'c}
\affiliation{Max Planck Institute for Astronomy (MPIA), K\"{o}nigstuhl 17, D-69117 Heidelberg, Germany}
\email{tomicic@mpia-hd.mpg.de}
\author{Annie Hughes}
\affiliation{Universit\'{e} de Toulouse, UPS-OMP, 31028 Toulouse, France}
\affiliation{CNRS, IRAP, Av. du Colonel Roche BP 44346, 31028 Toulouse cedex 4, France}
\author{Kathryn Kreckel}
\affiliation{Max Planck Institute for Astronomy (MPIA), K\"{o}nigstuhl 17, D-69117 Heidelberg, Germany}
\author{Florent Renaud}
\affiliation{Department of Astronomy and Theoretical Physics, Lund Observatory, Box 43, SE-221 00 Lund, Sweden}
\author{J\'{e}r\^{o}me Pety}
\affiliation{IRAM, 300 rue de la Piscine, F-38406 Saint Martin d'H\'{e}res, France}
\author{Eva Schinnerer}
\affiliation{Max Planck Institute for Astronomy (MPIA), K\"{o}nigstuhl 17, D-69117 Heidelberg, Germany}
\author{Toshiki Saito}
\affiliation{Max Planck Institute for Astronomy (MPIA), K\"{o}nigstuhl 17, D-69117 Heidelberg, Germany}
\author{Miguel Querejeta}
\affiliation{European Southern Observatory, Karl-Schwarzschild Strasse 2, D-85748 Garching bei M\"{u}nchen, Germany}
\affiliation{Observatorio Astron\'{o}mico Nacional (OAN), C/Alfonso XII 3, Madrid E-28014, Spain}
\author{Christopher M. Faesi}
\affiliation{Max Planck Institute for Astronomy (MPIA), K\"{o}nigstuhl 17, D-69117 Heidelberg, Germany}
\author{Santiago Garcia-Burillo }
\affiliation{Observatorio Astron\'{o}mico Nacional, Aptdo 1143, 28800 Alcal\'{a} de Henares, Spain}
\affiliation{Observatorio Astron\'{o}mico Nacional (OAN), C/Alfonso XII 3, Madrid E-28014, Spain}

\begin{abstract}

	We present the first spatially resolved ($\sim0.5$ kpc) measurements of the molecular gas depletion time $\rm\tau_{depl}$ across the disk of the interacting spiral galaxy NGC\,2276, a system with an asymmetric morphology in various SFR tracers.
    To estimate  $\rm\tau_{depl}$, we use new NOEMA observations of the $^{12}$CO(1-0) emission tracing the bulk molecular gas reservoir in NGC\,2276, and extinction-corrected H$\alpha$ measurements obtained with the PMAS/PPaK integral field unit for robust estimates of the SFR.
    We find a systematic decrease in $\rm\tau_{depl}$ of 1-1.5 dex across the disk of NGC\,2276, with a further, abrupt drop in $\rm\tau_{depl}$ of $\sim$1 dex along the galaxy's western edge.
    The global  $\rm\tau_{depl}$ in  NGC\,2776 is $\rm\tau_{depl}=0.55$ Gyr, consistent with literature measurements for the nearby galaxy population. 
    Such a large range in $\rm\tau_{depl}$ on sub-kpc scales has never previously been observed within an individual isolated or pre-merger system. 
 When using a metallicity-dependent molecular gas conversion factor X$\rm_{CO}$ the variation decreases by 0.5 dex.
    We attribute the variation in $\rm\tau_{depl}$ to the influence of galactic-scale tidal forces and ram pressure on NGC\,2276's molecular interstellar medium (ISM). 
    Our observations add to the growing body of numerical and observational evidence that galaxy-galaxy interactions significantly modify the molecular gas properties and star-forming activity within galactic disks throughout the interaction, and not just during the final merger phase.

\end{abstract}


\section{Introduction}\label{sec:intro}

 Star formation (SF) is a key process in the evolution of galaxies, affecting both their stellar populations and the properties of their  interstellar medium (ISM).
  The Star Formation Rate (SFR) and the bulk molecular gas (H$_2$) correlate well in nearby galaxies, both locally (e.g. \citealt{Bigiel08,Leroy13}) and globally (e.g. \citealt{Kennicutt98}).  
  The ratio between the H$_2$ mass  and SFR yields the depletion time of the H$_2$, i.e. the time needed to deplete the molecular gas reservoir assuming that the current SFR is constant, $\rm\tau_{depl}=M_{H_2}/SFR$.
  A characteristic  $\rm\tau_{depl}$ of 1-2\,Gyr is observed for local normal star-forming disk galaxies on the main-sequence (\citealt{Saintonge11,Leroy13}).
  Surveys of nearby galaxies show a scatter in $\rm\tau_{depl}$ of $\sim0.3$\,dex at galactic and sub-galactic scales (\citealt{Saintonge11,Leroy13}). 
  However, interacting  starburst  galaxies (\citealt{Klaas10,Nehlig16,Saito16}) and ultra-luminous infrared galaxies  (LIRGs, ULIRGS; \citealt{Saintonge11,Badenes12}) exhibit a lower systematic  $\rm\tau_{depl}$ of 0.05-0.8\,Gyr.  
  
  Investigations into the physics that drive variations in  $\rm\tau_{depl}$ among and within galaxies are still ongoing. 
  Stellar feedback and molecular cloud evolution have each been put forward to explain these variations, but there is increasing evidence that internal and external galactic  dynamics also affect $\rm\tau_{depl}$. 
  An example of internal dynamical processes is gravitational torques caused by galactic stellar structures, observed to modify the $\rm\tau_{depl}$ in the spiral arms of M51 (\citealt{Meidt13}). 
  Observations and numerical work indicate that external dynamical processes such as gravitation can also produce compressive and disruptive tides within galaxy gas disks during galaxy-galaxy interactions, leading to a broader distribution of $\rm\tau_{depl}$ (\citealt{Renaud14, Bournaud15}). 
  Ram pressure, as another external force, is known for quenching star formation, particularly in dwarf galaxies (\citealt{Steinhauser16}), but can also locally compress gas and have the opposite effect (\citealt{Ebeling14}), especially in more massive systems where the background potential helps slowing down gas stripping.
  Studies of $\rm\tau_{depl}$ in galaxies at various stages of interaction indicate that the tidal gravitational forces \textit{change}   $\rm\tau_{depl}$ up to 0.4\,dex \citep{Badenes12, Nehlig16, Lee17}. 
  \citet{Nehlig16} observed that ram pressure can  decrease  $\rm\tau_{depl}$, but not as effectively as the tidal effects.
  Within  starburst-like interacting galaxies, $\rm\tau_{depl}$ can vary by up to 1\,dex (\citealt{Saito16,Santaella16}).
 \citet[priv. comm.]{Renaud18inprep} also conclude  from their simulations of interacting galaxies that tidal forces generally decrease $\rm\tau_{depl}$ and increase its variation  within  galaxies. 
 The aforementioned studies only address moderate to late stages of galaxy interactions, where the galaxies are already colliding or interacting at small separation from each other.

 Here we study the spiral galaxy NGC\,2276, which is currently falling into NGC\,2300 group  and  interacting with the early-type galaxy NGC\,2300. 
 The NGC\,2300 group has four members including NGC\,2300 being the most massive one.
 Details about NGC\,2276 and the NGC\,2300 group are listed in Tab. \ref{tab:Tab01}. 
 NGC\,2276 itself exhibits high global SFR and an asymmetric distribution  in various multi-wavelength SFR tracers (X-Ray, FUV, H$\alpha$, infra-red and  radio; \citealt{Condon83,Gruendl93,Davis97,Rasmussen06}).
These different tracers indicate SFRs between 5-19.4\,M$_{\odot}$/yr (\citealt{Wolter15,Kennicutt83}). Thus for its stellar mass, NGC\,2276  is  too star-forming to be on the main sequence (expected SFR$\approx$5-6\,M$_{\odot}$/yr; \citealt{Elbaz07}).
NGC\,2276's total infra-red emission is $\rm\approx5.6\times10^{10}\,L_{\odot}$, which is not bright enough to be classified as a LIRG.

Previous papers (\citealt{Gruendl93,Hummel95,Rasmussen06,Wolter15}) argue that the enhanced and asymmetric SF in NGC\,2276 may be caused by tidal forces or ram pressure. 
While these papers argue that NGC\,2276 is in a phase after the first passage through the pericenter, they do not derive specific orbital characteristics for this system. 
  Tidal forces could be sufficient to trigger SF despite the large projected separation ($\approx75$\,kpc) to neighbor NGC\,2300, as \citet{Scudder12} show in their simulations that SFR may be enhanced by 0.3-0.6\,dex at large separations (up to 70\,kpc) between merging galaxies.  
 The presence of tidal forces in NGC\,2276 has also been invoked to explain the extended south-east arm in radio emission of NGC\,2276 (\citealt{Condon83}), and  truncation of the R-band continuum (\citealt{Gruendl93,Davis97}). 
Additional evidence for tidal forces includes a north-east extension in the I-band continuum of NGC\,2300 (\citealt{Forbes92,Davis97}), and the enhanced magnetic fields (\citealt{Hummel95}). 

 \begin{table*}[t!]
\centering
\caption{Global properties of NGC\,2276. }
\begin{tabular}{lll} \hline \hline
Parameter & Value & Reference \\
\hline 
RA  & $\rm07^{h}27^{m}13^{s}609$   &  Peak in $^{12}{\rm CO}(J=1\to0)$\\
DEC  & $\rm85^{d}45^{m}16^{s}361$  &  Peak in $^{12}{\rm CO}(J=1\to0)$\\
 Systematic velocity \big[km/s\big] & 2416  & Emission lines. \\
 Distance  \big[Mpc\big] & 35.5$\pm$2.5  & NED\footnote{\url{https://ned.ipac.caltech.edu/}}, \citet{Ackermann12} \\
 Scale \big[pc/arcsec\big] &  170$\pm$10 & \\
 Intergalactic medium density (IGM) $\rm\big[g\cdot cm^{-3}\big]$ & $\rm 10^{-27}$ & \citet{Mulchaey93,Rasmussen06} \\
 Projected distance from NGC 2300 \big[km/s\big] & 75$\pm$15  & \citet{Rasmussen06} \\
   $\rm\log M_{stellar}\big[M_{\odot}\big]$ of NGC\,2300   &  11.3  & From K-band, and from 3.4\,$\mu$m and 4.6\,$\mu$m (WISE) (using \citealt{Querejeta15}) \\  
 Line of sight velocity relative to IGM \big[km/s\big] & $\approx$300  & \citet{Rasmussen06} \\
 Inclination  & $\approx20^\circ\pm10^\circ$ & From radial velocities and radial stellar profile\\
 R$_{25}$ \big[kpc\big]  & 67$''=11.6$   & HyperLeda\footnote{\url{http://leda.univ-lyon1.fr/}}\\
 $\rm\log M_{stellar}\big[M_{\odot}\big]$    & 10.7$\pm$0.2   & From 3.4\,$\mu$m and 4.6\,$\mu$m (WISE) (using \citealt{Querejeta15}) \\
  $\rm\log M_{H_2}\big[M_{\odot}\big]$    & 9.8$\pm$0.05  & From $\rm^{12}{CO}(J=1\to0)$ estimated in this paper  \\
 $\rm\log M_{HI}\big[M_{\odot}\big]$    & 9.8   & \citet{Rasmussen06} \\
  $\rm M_{HI}/M_{stellar}$    & 0.13 &   \\
   $\rm\log L_{\rm IR}\big[L_{\odot}\big]$  &  $10.75\pm0.05 $  &  From IRAS; \citet{Sanders03}  \\
 $\rm  SFR(H\alpha,corr) \big[M_{\odot}/yr\big]$ &  17$\pm$5  & From the integrated spectra \\
  $\rm SFR(FUV+22\,\mu m) \big[M_{\odot}/yr\big]$ &  $\rm \approx10$ & From FUV and 22\,$\mu$m maps  \\
 $\rm  SFR \big[M_{\odot}/yr\big]$ &  $5-19$  & Literature (\citealt{Wolter15}, \citealt{Kennicutt83}).\\
\hline
\end{tabular}  \\ 
\label{tab:Tab01}
\end{table*}

 Enhanced X-ray emission outside NGC\,2276, and the bow-shock feature on the western edge of NGC\,2276's disk was attributed to ram pressure (\citealt{Rasmussen06}) as similar features have been observed in galaxies with ongoing ram pressure (\citealt{Paramo97,Sivanandam14,Troncoso16}). 
 The  high ram pressure acting on NGC\,2276 is linked to the unusually high density of the group's inter-galactic medium (\citealt{Mulchaey93}). 
 Simulations by \citet{Wolter15} show that ram pressure alone could explain the morphology and the lack of some HI gas in NGC\,2276.

Despite its exceptional SFR, the distribution of NGC\,2276's molecular gas reservoir has not previously been mapped at high spatial resolution. 
Spatial variations in $\rm\tau_{depl}$ could indicate if tidal forces and/or ram pressure have an impact on the ISM physics and $\rm\tau_{depl}$ as such in NGC\,2276.  
This letter presents observations of H$_2$ gas (as traced by CO emission) at sub-kpc scales and spatially resolved measurements of $\rm\tau_{depl}$ in NGC\,2276 for the first time. 
Additionally, we correct our IFU measurements of H$\alpha$ emission from the star-forming regions for internal extinction caused by dust, an important step that has not been applied to previous studies of SF in NGC\,2276 using narrowband imaging.

\section{Data}\label{sec:Data}

Observations with the integral field unit (IFU) PMAS in PPaK mode (\citealt{Kelz06}) on the Calar Alto 3.5m telescope are used to obtain spatially resolved H$\alpha$ emission.
We observed a mosaic of 6 pointings ($\approx75''$ in diameter) with three dither positions, covering the entire galaxy.  
The raw data were calibrated using the P3D software package (\citealt{Sandin10}) and established calibration procedures.
We used PanSTARRS images for astrometry and R-band images from the La Palma observatory (NED\footnote{\url{https://ned.ipac.caltech.edu/}}) for absolute flux calibration. 
The final data cube was re-sampled onto a grid with 1\,arcsec spatial pixels (spaxels) corresponding to $\approx170$\,pc.
The datacube is Nyquist-sampled with $\approx$3 spaxels across the instrumental point spread function. 
The reduced spectra have  a spectral resolution of R=1000 and cover 3700-7010\,\AA{}. 
We analyzed the reduced spectra  and extracted the emission lines using the GANDALF software package (\citealt{Sarzi06}). 
During the process, the spectra were  corrected for foreground Galactic extinction.  
The angular resolution of the final data is  2''.7 ($\approx450$\,pc). More details will be provided in \citet{Tomicic18inprep}.

\begin{figure*}[t!]
\centering
    \includegraphics[width=1.\textwidth]{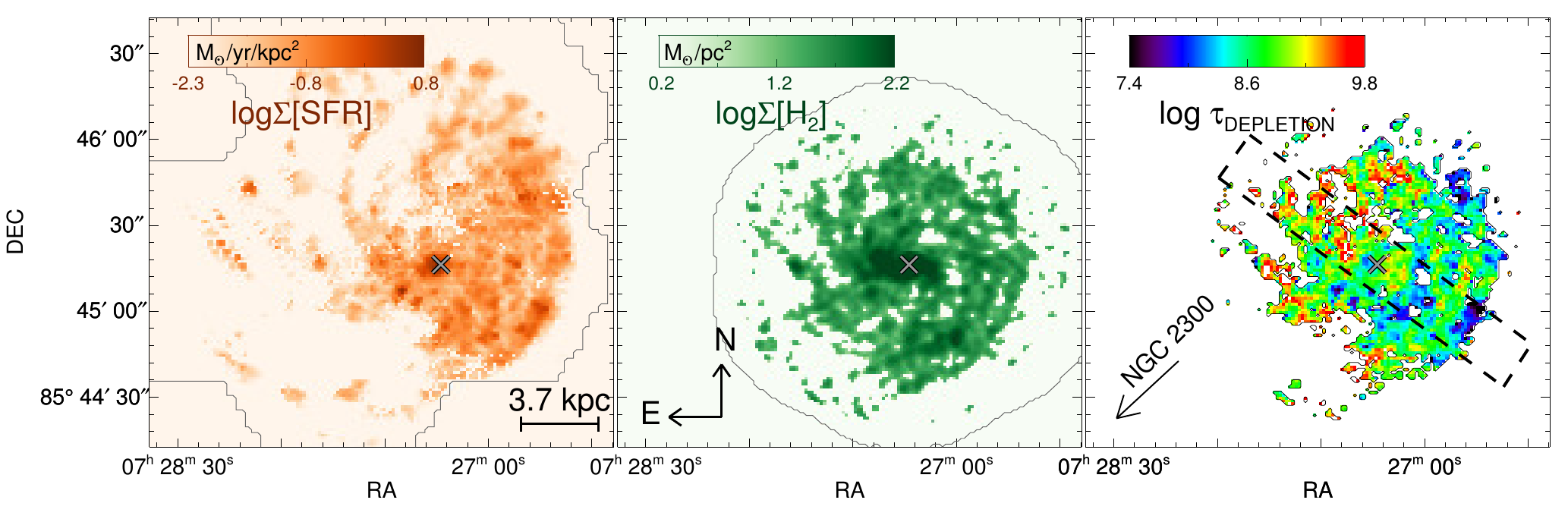}
    \caption{Distribution of $\rm\Sigma_{SFR}$(H$\alpha$,corr) (left), $\rm\Sigma_{H_2}$ (middle), and the depletion time $\rm\tau_{depl}$ ($=\Sigma_{H_2}/\Sigma_{SFR}$; right) across the disk of NGC\,2276.
    We indicate the areas observed in each tracers. 
    In the panel showing the $\rm\tau_{depl}$, we show the slit used to extract the data for the Kennicutt-Schmidt diagram in Fig.\,\ref{fig:Fig_2} by a dashed rectangle. 
    The slit orientation was chosen to encompass the largest range in $\rm\tau_{depl}$  values. Annotations indicate the direction towards the neighbor elliptical galaxy NGC\,2300 and  NGC\,2276's center (X symbol). }
    \label{fig:Fig_1}
\end{figure*}

\begin{figure*}[t!]
    \centering
    \includegraphics[width=1\textwidth]{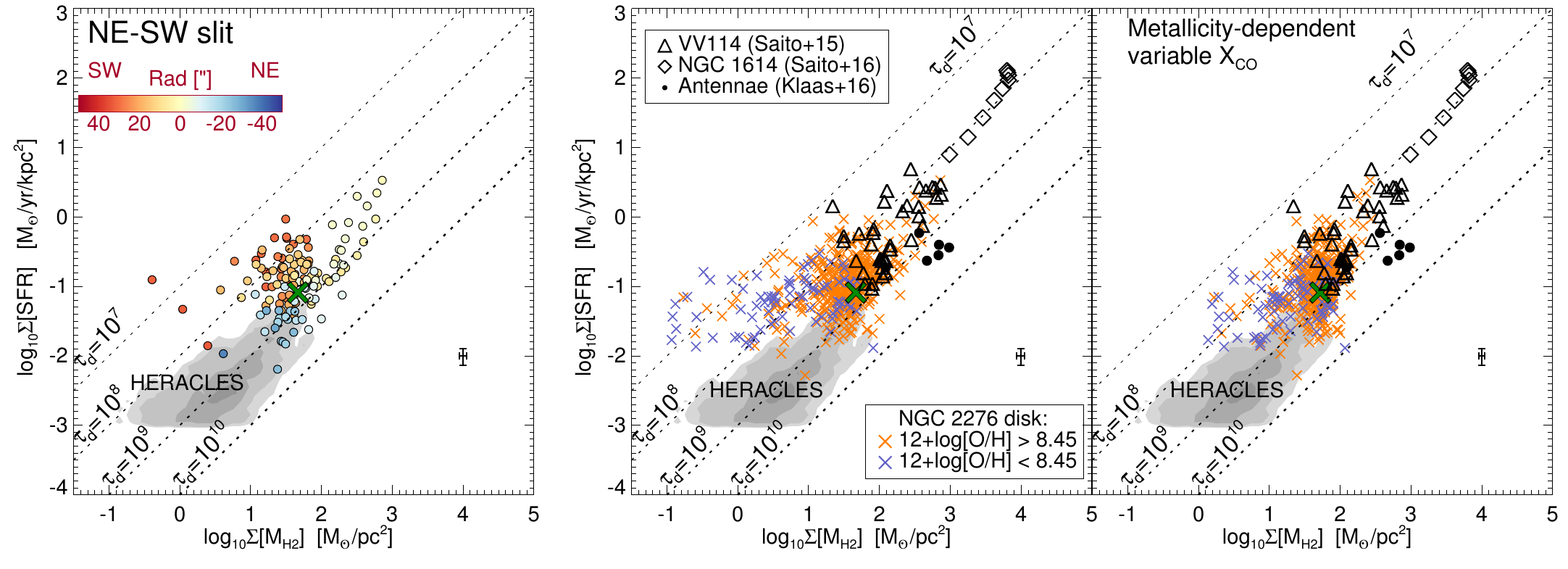}
    \caption{ Large variation in the $\rm\tau_{depl}$ is seen across different regions of NGC\,2276's disk, and presented in the $\rm\Sigma_{SFR}$(H$\alpha$,corr) vs. $\rm\Sigma_{H_2}$ diagram (Kennicutt-Schmidt diagram). The left panel shows NGC\,2276 data from the 20'' wide slit (shown in the right panel in Fig.\,\ref{fig:Fig_1}), which are color-coded from north-east (blue) toward south-west (red). The middle and right panels present data from NGC\,2276's entire disk (blue and orange crosses for different metallicity ranges), the mid-stage merger VV\,114 (\citealt{Saito15}), the luminous merger remnant NGC\,1614 (\citealt{Saito16}), and the Antennae (\citealt{Klaas10}). While we used constant X$\rm_{CO}$=$\rm2\times10^{20}cm^{-2} (K\cdot km/s)^{-1}$ for NGC\,2276 data in the left and middle panels, on the right panel we applied an X$\rm_{CO}$ factor that takes into account the spatial variation in nebular metallicity (\citealt{Narayanan12}).  The contours present the data from the HERACLES survey of nearby galaxies at sub-galactic scales (\citealt{Leroy13}), and the green X symbol is the mean galactic value for NGC\,2276 ($\langle\tau_{depl}\rangle$=0.55\,Gyr). The pixels from the NGC\,2276 maps are binned to sizes of 2.7'' ($\approx450$\,pc) to show spatially independent data. Typical uncertainties are shown in the right corner. The dashed lines indicate  $\rm\tau_{depl}$ of constant values.   }
    \label{fig:Fig_2}
\end{figure*}

To estimate the SFR surface density $\rm\Sigma_{SFR}$(H$\alpha$,corr), we use  extinction-corrected H$\alpha$ surface brightness $\Sigma$(H$\alpha$,corr). 
Based on  BPT diagrams (\citealt{Kewley06}) of emission lines, we find that the  H$\alpha$ emission arises from  star-forming regions and not from shocks.
 For the extinction correction, we assume the foreground screen model, apply the \citet{Cardelli89} extinction curve, assume H$\alpha$/H$\beta$=2.86 (case B recombination at a gas temperature of $\approx10^{4}$\,K) and a selective extinction R$_V=$3.1.
 To convert $\Sigma$(H$\alpha$,corr) to $\rm\Sigma_{SFR}$(H$\alpha$,corr), we use the SFR prescription  from \citet[][Eq. 1 and 2]{Murphy11}.
 We show $\rm\Sigma_{SFR}$(H$\alpha$,corr) map of NGC\,2276 in Fig.\,\ref{fig:Fig_1}.    

To estimate the mass surface density of the H$_2$ ($\rm\Sigma_{H_2}$), we mapped the $^{12}{\rm CO}(J=1\to0)$ emission from NGC\,2276 with the NOEMA interferometer at Plateau de Bure (NOrthern Extended Millimeter Array; project ID: w14cg001) and the IRAM 30m telescope (project ID: 246-14). The NOEMA observations consisted of a 19-point hexagonal mosaic (with a field of view 2.2' in diameter) centered on RA(J2000) $\rm07^{h}27^{m}14^{s}.55$ and Dec.(J2000) $\rm+85^{d}45^{m}16^{s}.3$. The 30m observations covered a $3\times3$ arcminute field centered on the same position. Both targeted the $\rm^{12}{CO}(J=1\to0)$ emission assuming a systemic LSR velocity of 2425\,km/s. The final combined (NOEMA+30m) cube has an angular resolution of 2.5''$\times$2.1'', a channel width of 5\,km/s, and $1\sigma$ sensitivity of 60\,mK per channel. For the analysis in this paper, we use a version of the cube that has been smoothed to 2.7'' resolution using a Gaussian convolution kernel. The sensitivity of this cube is 50\,mK per 5\,km/s channel. More details will be presented in \citet{Hughes18inprep}.  

For $\rm\Sigma_{H_2}$, we assumed the Galactic value X$\rm_{CO}$=$\rm2\times10^{20}cm^{-2}\,(K\cdot km/s)^{-1}$ (\citealt{Bolatto13}) of the conversion factor. 
We show the $\rm\Sigma_{H_2}$ map of NGC\,2276 in Fig.\,\ref{fig:Fig_1}. 
We use this conversion factor as NGC\,2276's nebular metallicity, estimated from the [NII]/[SII] and [NII]/H$\alpha$ ratios and using Eq. 3 in \citet{Dopita16}, is similar to the solar value (log[O/H]+12 ranges between 8.4 and 8.9). 
We also present in Fig. \ref{fig:Fig_2}  the NGC\,2276 data for the case of a spatially varying X$\rm_{CO}$ factor taking into account local variation in metallicity.

\section{Results} 

\subsection{The depletion time}\label{sec:Results_Depl}

The $\rm\Sigma(H_2)$ distribution is consistent with a fairly normal disk while $\rm\Sigma_{SFR}$(H$\alpha$,corr) show a prominent asymmetry toward the western edge (Fig.\,\ref{fig:Fig_1}). 
The resulting $\rm\tau_{depl}$ distribution is shown in Fig.\,\ref{fig:Fig_1}.
The standard deviation of $\rm\tau_{depl}$ values is 0.52 dex.
 The highest observed $\rm\tau_{depl}$(H$_2$) value is 9 Gyr, and it  gradually decreases to 0.1 Gyr  across the disk, from  north-east (NE) to south-west (SW). 
 The lowest $\rm\tau_{depl}$ values (10\,Myr-100\,Myr) are found along the western edge of the disk. 
 The mean galactic $\rm\tau_{depl}$ value is 0.55\,Gyr.
 From the integrated spectra, we estimate NGC\,2276's galactic SFR to be  
 $\approx$17$\pm$5\,M$_{\odot}$/yr.  

To demonstrate the amplitude of the variation in $\rm\tau_{depl}$ in NGC\,2276, we plot the pixel-by-pixel data on the Kennicutt-Schmidt diagram (Fig.\,\ref{fig:Fig_2}). 
The left panel shows NGC\,2276 data from the 20$''$ wide slit oriented in NE-SW direction (that follows the $\rm\tau_{depl}$ gradient), and other panels present NGC\,2276 data from the entire disk. 
The right panel shows NGC\,2276 data from the entire disk where we used a variable  X$\rm_{CO}$ factor corrected for local variation in metallicity (\citealt{Narayanan12}). 
The contours show the data from the HERACLES survey of nearby galaxies (\citealt{Leroy13}), and $X$ symbol represents the NGC\,2276's mean galactic value. 
The HERACLES survey examines $\sim$1 kpc regions in 30 galaxies. 
We added sub-galactic regions from the mid-stage merger VV\,114 (\citealt{Saito15}), luminous merger remnant NGC\,1614 (\citealt{Saito16}), and Antennae  (\citealt{Klaas10}). 
The NGC\,2276 data from the slit show a decrease in $\rm\tau_{depl}$ from 3\,Gyr to 10\,Myr from NE toward SW, while the center exhibits a $\rm\tau_{depl}$ of about 0.4\,Gyr. 
The $\rm\tau_{depl}$ values in the disk show a $\approx$0.5\,dex narrower range when we use metallicity-dependent X$\rm_{CO}$ factor compared to when we use a single X$\rm_{CO}$ factor.
 The change in $\rm\tau_{depl}$ is most pronounced in the outskirts of the disk, esp. the Western edge, where metallicities are lower.
 However, we caution that metallicity estimates in the Western edge region could potentially be affected by the stellar cluster's age, and thus ionization parameters (see Sec. \ref{sec:Results_Tidal_Ram}).

\begin{figure}[t!]
    \centering
    \includegraphics[width=0.5\textwidth]{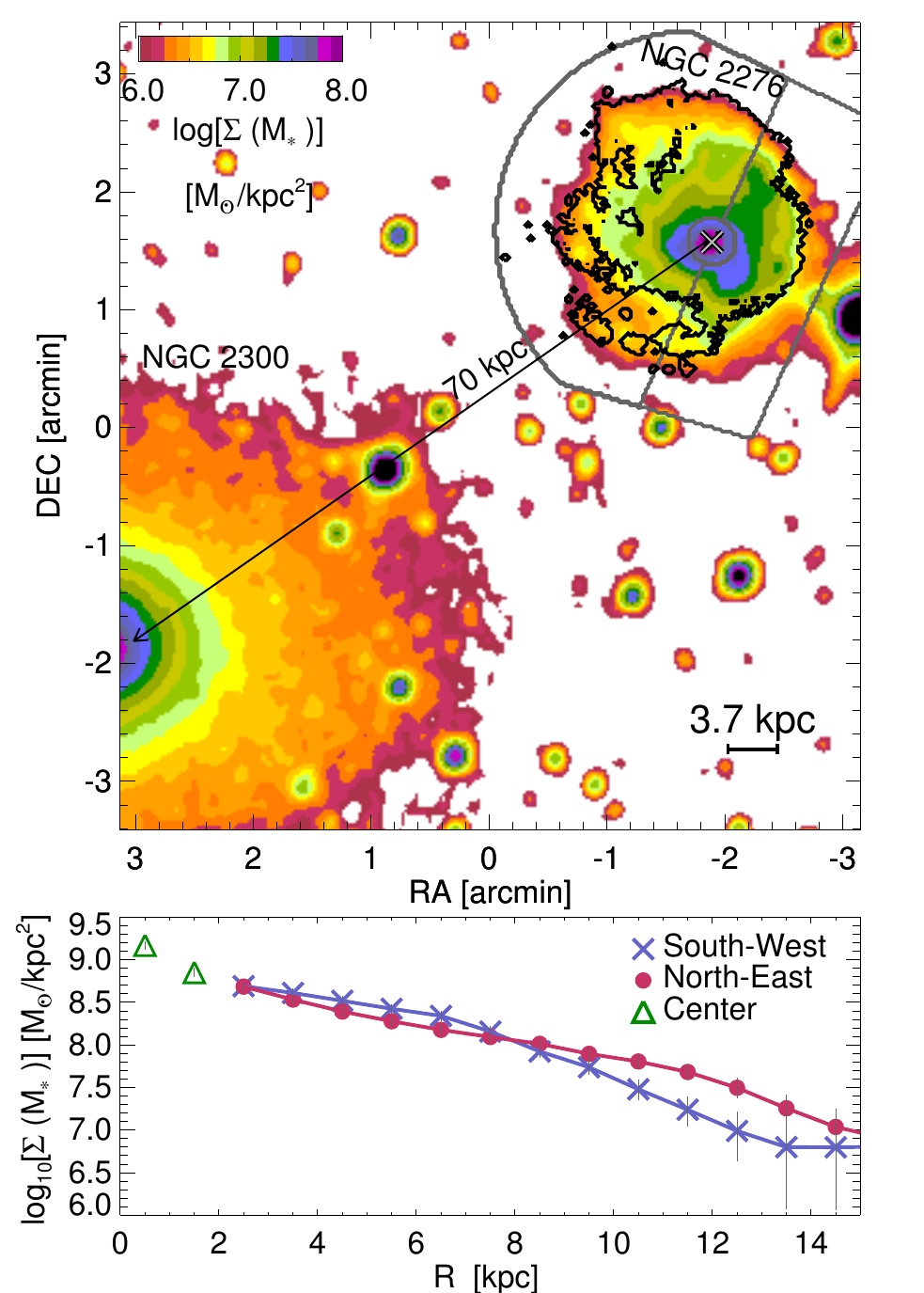}
     \caption{$\rm\Sigma(M_{stellar})$ of NGC\,2276 and NGC\,2300 derived from WISE images. NGC\,2276's stellar mass distribution is asymmetric and has a shorter scale-length on the south-west side compared to the north-east side. We attribute this to tidal forces exerted by NGC\,2300. The black contour on NGC\,2276 shows the observed H$\alpha$ emission. The projected distance between the galaxies is marked. We show below a radial profile of $\rm\Sigma(M_{stellar})$ for the south-west (crosses), central (triangles), and north-east (circles) sides of NGC\,2276 that are marked on the upper panel }
    \label{fig:Fig_3}
\end{figure}

\begin{figure}[t!]
    \centering
    \includegraphics[width=0.5\textwidth]{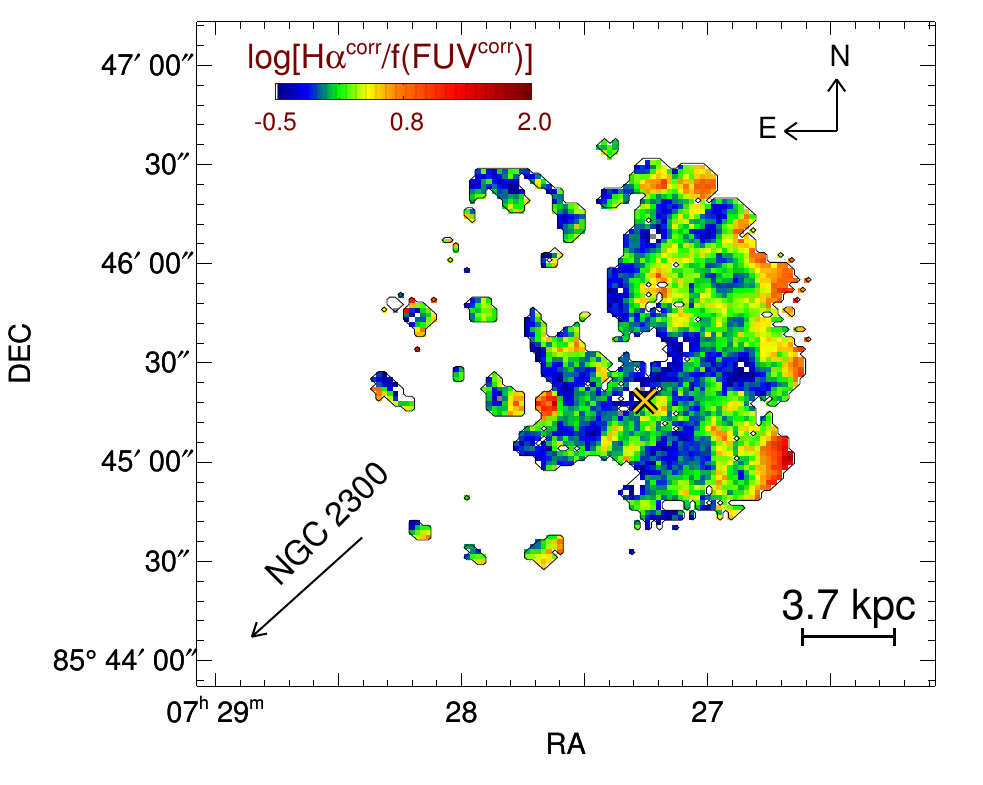}
     \caption{H$\alpha$,corr/f$_\nu$(FUV,corr) ratio map, with lower ratios robustly indicating older clusters (\citealt{SanchezGill11}). Shown data have S/N$>5$ in both Balmer lines and FUV emission. The mean uncertainty of the data shown is $\approx0.3$\,dex. We attribute the increase in the ratio on the western edge of the disk to ram pressure triggering recent star formation.}
    \label{fig:Fig_4}
\end{figure}

\subsection{Tidal forces and ram pressure}\label{sec:Results_Tidal_Ram}
  
 Galactic-scale tidal forces are responsible for features such as stellar streams, disk thickening and asymmetries in stellar disks. 
 We derived $\rm\Sigma(M_{stellar})$ map of NGC\,2276 and NGC\,2300 from WISE images at 3.4\,$\mu$m and 4.6\,$\mu$m following Eq. 8 in \citet{Querejeta15}.
 The resulting map on Fig.\,\ref{fig:Fig_3}, confirms that the $\rm\Sigma(M_{stellar})$ distribution in NGC2276 is strongly asymmetric, and shows a steeper drop on the SW side compared to the NE side. While other external (e.g. minor mergers, gas accretion) or internal (asymmetries in the dark matter halo) mechanisms cannot be ruled out as the origin of these features (\citealt{Laine14}), we propose (as previous authors have done) that the asymmetric $\rm\Sigma(M_{stellar})$ in NGC\,2276 is due to tidal forces.

To compare NGC\,2276 to other galaxies, we quantify the tidal strength of the interaction $Q$ experienced by NGC\,2276 following Eq. 1 in \citet{ArgudoFernandez15}, i.e.
\begin{equation}
Q = \log_{10}\left[\frac{M_{comp}}{M_{2276}}\left(\frac{D_{25}}{r}\right)^{3}\right]\ ,
\end{equation}
where $\rm log(M_{comp}/M_{\odot})=10^{11.3}$ and $\rm log(M_{2276}/M_{\odot})=10^{10.7}$ are the stellar masses of NGC\,2300 and NGC\,2276, respectively; $D_{25}$ is the B-band optical diameter of NGC\,2276, and $r=75$\,kpc is the projected separation 
between NGC\,2300 and NGC\,2276. 
For NGC\,2276, we find $Q=-0.9$, which is significantly higher than the typical value for isolated galaxies ($Q=-5.2\pm0.8$) and on the high end of isolated galaxy pairs ($Q=-2.3\pm1.2$, \citealt{ArgudoFernandez15}).

 As well as galactic-scale tides, our new observations also show evidences for ram pressure affecting NGC\,2276. 
 First, the scale-length of the  ionized gas on the SW side of NGC\,2276's disk is significantly shorter (up to 1-2\,kpc) than the stellar emission scale-length (Fig.\,\ref{fig:Fig_3}). 
 In contrast, the ionized gas follows well the stellar distribution on the NE side. 
 This feature cannot be explained by tidal forces alone, and may be a signature of ram pressure stripping of the interstellar gas. 
 Secondly, the H$\alpha$,corr/f$_\nu$(FUV,corr) ratio increase along the western rim of NGC\,2276's disk (Fig.\,\ref{fig:Fig_4}). 
 We retrieved the FUV images from the public AIS survey\footnote{\url{http://galex.stsci.edu/GR6/?page=tilelist&survey=ais}} (\citealt{Bianchi14}). 
 To calibrate the FUV images, we subtracted the background emission from NGC\,2276, and corrected the FUV map for the foreground Milky Way extinction (applying E$\rm_{B-V}=0.088$).
 The H$\alpha$,corr/f$_\nu$(FUV,corr) ratio robustly indicates the age of stellar clusters  (\citealt{SanchezGill11}), showing that the westernmost regions are dominated by the youngest clusters.
 We link this most recent SF on the western edge of the disk to ram pressure (as similarly observed in the Large Magellanic Cloud by \citealt{Piatti18}).

\section{Discussion and summary}\label{sec:Discussion}

In this letter, we have presented spatially resolved measurements of the H$_2$ and $\rm\tau_{depl}$ in NGC\,2276 for the first time. 
On galactic scales, the mean $\rm\tau_{depl}$ of NGC\,2276 is 0.55\,Gyr, which is lower than the $\rm\tau_{depl}$=1-2\,Gyr found in surveys of nearby galaxies (COLD GASS, HERACLES; \citealt{Saintonge11,Leroy13}), but still within the $\rm\tau_{depl}$ range of those galaxies (Fig.\,\ref{fig:Fig_2} or Fig.\,14 in \citealt{Leroy13}). 
We note that  NGC\,2276 exhibits  $\rm\Sigma_{SFR}$(H$\alpha$,corr) and $\rm\Sigma_{H_2} $ values that are higher than in the HERACLES survey, and lower values than in the galaxies  at the coalescence phase (\citealt{Saito15,Saito16}). 
On the other hand, we observe a large variation  in  $\rm\tau_{depl}$ at sub-galactic scales in NGC\,2276. 
On a pixel-to-pixel scale (pixels $\approx450$\,pc in size) in a 20'' wide NE-SW slit, $\rm\tau_{depl}$ ranges from 10\,Myr to 3\,Gyr. This is almost \textit{2-3 orders of magnitude variation} in $\rm\tau_{depl}$ within a single disk.
 Furthermore, our results reveal a \textit{gradual decrease} in $\rm\tau_{depl}$ across the disk in the NE-SW direction.   
 
 This is a factor of $\approx$30 larger range at sub-galactic scales compared to other nearby galaxies. 
 For individual galaxies in the HERACLES survey, sub-galactic regions show a typical spread of $\approx$0.5\,dex (Fig.\,18 and 19 in \citealt{Leroy13}). 
 However, a spread in NGC\,2276's $\rm\tau_{depl}$ decrease by 0.5\,dex (down to 2.5\,dex) when we use variable metallicity-dependent X$\rm_{CO}$ factor, which indicates that sub-galactic variation in $\rm\tau_{depl}$ may be affected by a different metallicity prescriptions or when using a single  X$\rm_{CO}$ factor. 
 The NGC\,2276's variation in the $\rm\tau_{depl}$  is comparable only with the merging starburst LIRGs observed by  \citet{Santaella16} and \citet{Saito16}, although their mean galactic values exhibit lower $\rm\tau_{depl}$ than NGC\,2276. 
 The mid-stage merging galaxy VV114 (\citealt{Saito16}) covers a similar range in parameters ($\rm\Sigma_{SFR}(H\alpha,corr)$ and $\rm\Sigma_{H_2}$) as NGC\,2276 on the Kennicutt-Schmidt diagram and shows almost 2\,dex variation in $\rm\tau_{depl}$.
 \citet{Renaud18inprep} find a 1-3\,dex difference in $\rm\tau_{depl}$ between regions in their simulations of the Antennae  during early phases of interaction.  
 However, the observed variation in $\rm\tau_{depl}$ is only 0.5\,dex in late-phase merging LIRGs such as the Antennae (\citealt{Klaas10}) and NGC\,$4567/8$ (\citealt{Nehlig16}).

 Based on the clear asymmetric distribution of the stellar disk, we tentatively attribute the large-scale gradient in $\rm\tau_{depl}$ as to tidal forces acting on NGC\,2276. 
 The tidal forces act on the entire disk, and likely cause a gradual 1-1.5\,dex  decrease  of $\rm\tau_{depl}$ between the two sides of the disk. 
 The ram pressure further disturbs the morphology of the gas disk, and particularly compresses gas on its western edge, which has younger stellar clusters and 1\,dex lower $\rm\tau_{depl}$ compared to the rest of the disk.

 NGC\,2276 shows that galaxies in the pre-coalescence phase of interaction may already exhibit large variations in $\rm\tau_{depl}$ at sub-galactic scales, while still showing a typical $\rm\tau_{depl}$ value for the galaxy-wide average. Our observations demonstrate clearly that a galaxy-galaxy interaction significantly modifies the star formation efficiency of molecular gas locally, that the effect is distributed throughout the galactic disk and not just at the galaxy center, and that these changes occur well before coalescence.

\textit{Facilities}: IRAM (NOEMA and 30m), CAHA (PMAS) \\
\textit{Software}: P3D (\citealt{Sandin10}), GANDALF (\citealt{Sarzi06}) \\

\acknowledgments
We thank the referee for constructive comments that improved this letter.  
We thank M\'{o}nica Rela\~{n}o, Rebecca McElroy and Sharon Meidt for constructive comments on the paper. NT and KK acknowledge grants SCHI 536/8-2 and KR 4598/1-2 from the DFG Priority Program 1573. FR acknowledges support from the Knut and Alice Wallenberg Foundation. JP acknowledges support from the Program National "Physique et Chimie du Milieu Interstellaire" (PCMI) of CNRS/INSU with INC/INP, co-funded by CEA and CNES. This work is based on observations made with the NASA Galaxy Evolution Explorer (GALEX). GALEX is operated for NASA by the California Institute of Technology under NASA contract NAS5-98034. This work is based on observations carried out under project number w14cg001 with the IRAM NOEMA Interferometer and 30m telescope. IRAM is supported by INSU/CNRS (France), MPG (Germany) and IGN (Spain). This work is also based on observations collected at the Centro Astron\'{o}mico Hispano-Alem\'{a}n (CAHA), operated jointly by the Max-Planck Institut f\"{u}r Astronomie and the  Instituto de Astrof\'{i}sica de Andaluc\'{i}a (CSIC).

%\bibliographystyle{aasjournal}%\bibliographystyle{yahapj}
%\bibliography{}

\end{document}